# An ab initio Local Exchange Approach to the Calculations of $T_c$ of Dilute Magnetic Semiconductors

V.G. Yarzhemsky[1,2,*], S.V. Murashov[1] and A.D. Izotov[1]

1) Kurnakov Institute of General and Inorganic Chemistry, 119991, Moscow Leninsky 31, Russian Federation

2) Moscow Institute of Physics and Technology, 141701, Dolgoprudny, Institutsky Lane 9, Moscow Region Russian Federation

ABSTRACT

The DFT calculations were performed of densities of states of GaAs and $Ga_{0.9375}Mn_{0.0625}As$. It is obtained that a part of Mn$3d$- states is hybridized with valence band at Fermi level. The exchange integrals of Anderson impurity model were calculated making use of atomic Hartree-Fock package and angular momentum coupling technique. Theoretical $T_c$ of $Ga_{0.9375}Mn_{0.0625}As$ obtained in the multiscale ab initio method is in reasonable agreement with experiment. The application of Hubbard parameters in DFT calculations using ultrasoft pseudopotentials is discussed.

___________________________________

[*] Corresponding author e-mail vgyar@igic.ras.ru

INTRODUCTION

Magnetic semiconductors are promising materials for spintronics, i.e electronics, processing and recording information involving electron spin in addition to its charge[1]. The promising way to obtain these materials is adding of magnetic $3d$ elements into known semiconductors, e.g .GaAs, InSb, $CdGeAs_2$ [1-4]. Unfortunately the crystal structures of magnetic materials differ from that of semiconductors and solubility of magnetic atoms in these semiconductors is several units of percent only, and these materials are often called DMS (dilute magnetic semiconductors) The highest substitution rate of Ga in GaAs by Mn is about 0,053 and Curie temperatute is 110K [1]. Despite the fact that the composition, properties and electronic structures of DMS are known, there is still no consensus about the type of the exchange interaction, determining a transition to the ferromagnetic state. Several years ago Ohno wrote, that the understanding of the ferromagnetism of (Ga,Mn)As is not adequate, however [1]. Until now this assertion seems to be true.

Photoemission and optical studies of DMS demonstrated that Mn provides both localized spins and itinerant holes mutually coupled by an exchange interaction. Zener first proposed the model of ferromagnetism driven by the $p–d$ exchange interaction between band carriers and localized spins [5]. Present theoretical models are closely connected with the relative energy positions of Fermi level, valence band, and so-called impurity band , i.e. $3d$- electrons of magnetic impurity [6]. In a double exchange model $d$- level of transition metal is located in the band gap for the spin-up states. In a $p$-$d$ exchange model the majority spin $d$- level lies below the valence $p$- band and the a minority-spin level above it [6]. The results of channelling experiments, which measure the concentrations both of Mn ions and of holes relevant to the ferromagnetic order, with magnetization, transport, and magneto-optical data provide strong evidence, that the location of the Fermi level in the impurity band determines $T_C$ through determining the degree of hole localization [7]. Hard x-ray photoelectron investigations of $Ga_{1-x}Mn_xAs$ strongly favor a model in which there is no gap between the Mn-induced impurity band and the GaAs valence band, and suggest that the magnetism originates from the coexistence of the two different mechanisms discussed above, i.e. double exchange and $p$-$d$ exchange [8]. Resonant photoemission experiments in agreement with theoretical calculations of $Ga_xMn_{1-x}As$ resulted in the position the Mn$3d$- states primarily in the valence band of the GaAs host band, with the Mn$3d$- states extending over a broad range of energies. The experimental spectrum has a main peak of $3d$-electrons at 3.2 eV binding energy [9]. The ARPES results unambiguously demonstrate that the Fermi level resides deeply inside the As-$4p$ valence band [10]. However making use of ARPES measurement in soft x-ray region it was concluded that experimental dispersions of the host GaAs band around the Γ point indicate that the valence band maximum of the host GaAs band stays below $E_F$ (Fermi energy) [11]. The infrared optical data proved that $E_F$ resides in a Mn-induced impurity band, but cannot exclude its overlap with the valence band in the density of states. [12]. These data were interpreted in impurity band model. On the other hand, optical spectroscopy measurements in a large set of systematically prepared (Ga,Mn)As epilayers [13] show that doping trends [7] in a limited number of samples are not generic and not even prevailing in ferromagnetic (Ga,Mn)As materials. The key difference between the Zener model [5] and Anderson impurity-band approach [14] is the nature of the wavefunctions that mediate ferromagnetism: in the former, the wavefunctions take on the character of the GaAs host semiconductor, whereas in the latter, they always remain impurity like, no matter the Mn concentration. Some contradictions of experimental results on relative positions of valence band of host semiconductor and so called impurity band and discussion of this problem was even called "battle of band" [15]. On the other hand, the essential role of the careful optimization of (Ga,Mn)As synthesis was demonstrated and it was concluded, that by recognizing that the bands are merged, that is, overlapped and mixed, in the optimized ferromagnetic (Ga,Mn)As materials, the distinction between "valence" and "impurity" bands becomes mere semantics with no fundamental physical relevance [13].

In addition to the position of the impurity band, recent spectroscopic investigations revealed some other interesting featured of DMS electron structure. According to the spectra of magnetic circular dichroism *sp-d* exchange interaction in ferromagnetic $Ga_{0.97}Mn_{0.03}$ is localized [16]. Zeeman splitting at Γ point is 120 meV, but at *L* point it is 8 meV only [16]. Resonant tunneling spectroscopy, applied to a variety of surface GaMnAs layers showed that the valence band structure of GaAs does not merge with the impurity band and the exchange splitting of the valence band is found to be very small (only several meV), even in GaMnAs with a high $T_c$ (154 K) [17]. Recent photoelectron spectroscopy data [18] showed a highly dispersive Mn-induced energy band above the valence band maximum of the host material. The development of this band can be observed at Mn concentrations below 0.5%. For concentrations above 1%, this band reaches the Fermi level (that is located in the band gap of GaAs) and can host holes mediating the ferromagnetism [18]. Resonant tunneling experiments of GaMnAs showed that at low Mn concentrations the dispersionless impurity band lies in the band gap of the host semiconductor [19]. When the Mn concentration is close to that, corresponding to ferromagnetic transition the bands merge, but at higher Mn concentrations initial band ordering is restored [19]. The dependence of Mn3*d*- states localization relative to the host band on Mn concentration in $Cd_{1-x}M_xGeAs_2$ was obtained theoretically [20]. Theoretical calculations using hybrid HSE06 functional [21] resulted in the electron structure of $Ga_{1-x}Mn_xAs$ in agreement with photoelectron spectra and pressure dependent. It was concluded that Mn-derived spin-polarized feature in the majority spin band gap is not detached from the host valence band [21].

The first-principle calculation of $T_c$ in DMS are usually based on mean field approximation  DFT LDA+U approach [6,21,22]. Anderson impurity model takes into account exchange interaction localized on Mn center. Anderson wrote : the Hartree-Fock fields for electrons of different spins differ not only by exchange integrals but by true Coulomb integrals, and only this circumstance makes localized moments possible in the  iron group. The Anderson *s-d* interaction term was written as:

$$H_{sd} = \sum_{k,\sigma} V_{dk} \left( c_{k\sigma}^* c_{d\sigma} + c_{d\sigma}^* c_{s\sigma} \right) \qquad (1)$$

This type of *s-d* interaction is a purely one-electron energy, entirely different from the *s-d* exchange interaction which enters the Zener type of theory [14]. The on site *s-d* interaction contribution to magnetic ordering was calculated in our previous work making use multiscale HF-DFT (Hartree-Fock and Density Functional Theory) method [23-24]. The resulting values of $T_c$ (Curie temperature) look underestimated and in present work the approach is extended in order to take into account exact atomic exchange integrals *s-d* and *s-p* and *d-d* Coulomb integrals. Note that whilst the *d-d* one center integrals of the 2$^{nd}$ and of the 4$^{th}$ order are called Coulomb, their function is the same as that of exchange *s-d* and *p-d* integrals, i.e. they minimize the energy of high spin state. The on site parameters, called Hubbard energies are included in modern solid state DFT packages [25-27], which use ultrasoft pseudopotentials [28]. The on-site Hubbard parameters are usually used to take into account large on-site Coulomb interaction. In the present work we take into account that ultrasoft pseudopotentials correspond well to exact atomic wavefunctions (see Figure 1 in Ref. [28]), and reconsider the choice of Hubbard parameters making use of atomic Hartree-Fock wavefunctions.

METHOD OF CALCULATION AND RESULTS

According to the Anderson impurity model [14], the electrons of the valence band in the vicinity of the magnetic atom can be considered as atomic. In the case of charge 2+, the number of Mn3*d* electrons does not change and the ground state is Mn $3d^5(^6S)$. In the compounds under consideration, the 3*d*-element atoms occupy the positions of trivalent Ga atoms and Mn ground state is $3d^4(^5D)$. In the present work calculations of electron structure of the host semiconductor and DMS were performed within DFT GGA (generalized gradient approximation) [25]. Theoretical DOS (densities of states) of the host semiconductor GaAs are presented in Figure 1. It is seen on Figure 1, that the valence band of the host semiconductor consists of *p*-orbitals of Ga and As. When the Ga is replaced by Mn in 1/16 of the formula units, a contribution of Mn3*d* and Mn4*s* orbitals to the valence band appears (see Figure 2). Near Mn atom, due to the exchange interaction at one center, the energy of the *s*- electrons with a spin parallel to the $3d^5$ spins (this state is denoted as $3d^5(^6S)4s(^7S)$) is below the energy of the state in which the spin of the *s*-electrons is antiparallel to the $3d^5$ spins, i.e. the state is $3d^5(^6S)4s(^5S)$. In the absence of magnetic ordering, the probabilities of spin-up and spin- down states are proportional to their statistical weights and this average state is denoted as $3d^5(^6S)4s_{av}$. In the model under consideration, the energy difference between the terms $3d^5(^6S)4s_{av}$ and $3d^5(^6S)4s(^7S)$ multiplied by the contribution $c_s$ of Mn4*s* electrons to the valence band equals to the energy of ferromagnetic ordering. In the case of filled valence band total spin is zero. In this case the spin-up polarization of *s*- electrons on Mn is accompanied by the same spin-down polarization of *p*-electrons on Mn. The energy gain on magnetic center is proportional to the difference of exchange integrals of *d*- core electrons with *s*- and *p*- electrons of valence band. After ferromagnetic phase transitions this energy gain is attained at all centers. When the Mn atom is placed at trivalent Ga site according to general chemical rules about one *d*- electron takes part in chemical bonding. This statement is confirmed by the results of calculation [6]. The energy of valence band electron in spin-up state is also less than in spin-down state. The redistribution of *d*- and *p*- electrons at Mn site between spin-up and spin-down states also results some energy gain. The model is illustrated in Figure 3. In paramagnetic case *s*-, *p*- and *d*- electrons on Mn equally populate spin-up and spin-down subbands. In ferromagnetic case spin-up subband is mostly populated by Mn4*s* and Mn3*d* electrons. At the same time population of spin-down subband by Mn4*p* electrons increases. The energy gain is due to the difference of exchange (Coulomb) integrals between localized 3*d*- electrons and valence band *s*-, *p*- and *d*- electron. At other atoms only *p*-band is populated. In this case exchange energy can be changed only due appearance of holes in a valence band or electrons in conduction band. Thus the important feature of the present model is that the filled valence band remains spin unpolarised and the energy gain is achieved due to redistribution of *s*-, *p*- and *d*-electron density between spin-up and spin-down subbands. The low spin-polarization of valence band of the present model band is in agreement with experimental data [16,17].

This multiscale model requires exact calculation of on-site integrals and electron structure of the material. Atomic wavefunctions were calculated making use of atomic Hartree-Fock package [29]. The ultrasoft potentials used in the calculation of solids [28] represent well the average configuration. Hence the energy of an electron in atomic field should be calculated relative to the average term. In this case the energy of *nl*- electron coupled to atomic term of 3*d*-shell may be represented as follows:

$$U_l = \sum_\kappa (3d, nl | R^\kappa | nl, 3d) \overline{C}^\kappa_{LS,L'S'}(3d, nl) \qquad (2)$$

The first multiplier is a Coulomb exchange integral of multipolarity κ and the second term is a coefficient, calculated making use angular momentum coupling technique [30,31] relative to the average term. This coefficient depends on the atomic term $LS$ and on the term $L'S'$ of atom with an extra *s*- or *p*- electron. The overline denotes that the coefficients are calculated relative to the average term. The sum includes one item κ=2 for *s* electrons and two items κ=1,3 for *p*-electrons.

Since in a filled band the spin-up polarization of *s*-electrons is accompanied by spin-down polarisation of *p*-electron, the energy gain per one Mn center may be represented as:

$$\Delta E = c_s (U_s - U_p) \qquad (3)$$

For the *d*- electron the Hubbard energy is calculated according to formula:

$$U_d = \frac{1}{n} \sum_\kappa (3d, 3d | R^\kappa | 3d, 3d) \overline{C}^\kappa_{LS}(3d, 3d) \qquad (4)$$

Where the sum includes κ=2,4 and represents the energy of ground term of $d^n$ configuration and *n* is the number of *d*- electrons.

This formula represents additional electron energy shift of atomic term (per one electron) relative to the energy in an average field.

A single-center exchange integrals, the coefficients and Hubbard energies $U_l$ obtained on their basis of are given in Table 1. It should be noted that since we consider spin-up terms relative to the average term, all these extra energies are negative. This assertion is in agreement with Hund's rule and with an assumption, that ultrasoft potentials exactly correspond to the average term.

DISCUSSION OF THE RESULTS

It is seen from Figure 2, which shows the DOS for the $Ga_{0.9375}Mn_{0.0625}$, that the introduction of Mn leads to a hybridization of the Mn3*d* with valence band in broad energy region. The position of the DOS maximum for d-states, which is -2.7 eV (relative to the Fermi level) is in good agreement with the experimental value of -3.2 eV, obtained by resonant photoemission [9] and other DFT calculations [9,21]. Small maximum with energy of -0.3 eV is also observed in the spectrum of resonant photoemission [9].

Total energies were calculated without Hubbard energy parameter and with and Mn4*s* Hubbard energy parameter -0,248 eV, i.e. the difference of 4*s* and 4*p* exchange energies with $3d^5(^6S)$ state for high spin state relative to the average term. The difference of total energies per one formula unit for this case is considered as energy of magnetic ordering.

To calculate the Curie temperature $T_c$ the formula [6] was used:

$$k_B T_c = \frac{2}{3}\Delta\varepsilon \qquad (5)$$

Where $\Delta\varepsilon$ is the energy per one formula unit. Contributions of $s$-$d$ exchange presented in Table 1 are significantly less, than experimental $T_c$=110 K [1]. Similar calculations with Hubbard energies for 3$d$- electrons are not possible, since only a small part of all 3$d$- DOS contributes to the valence band. Figure 2 shows, that $d$- DOS split into two parts, separated by a deep minimum. In our model we assumed that small part in the energy region close to $E_F$ contributes to valence band. This part near Fermi level is shown in the inset of Figure 2. The integral of $d$- DOS in the interval from the deep to $E_F$ equals to 0.265. This value was used in the following formula to calculate the energy of magnetic ordering due to spins-up orientation of $d$-electrons in the valence band:

$$\Delta E = c_d (U_d - U_p) \qquad (6)$$

Theoretical $T_c$ which includes spin polarization of $d$- and $s$- electrons in valence band equals to 166 K. This value is close to experimental values 110 [1] K and 188 K [13].

CONCLUSION

Our multiscale DFT and atomic Hartree-Fock calculation of $T_c$ in Ga$_{0.9375}$Mn$_{0.0625}$As resulted in reasonable agreement with experimental values. This is an ab initio version of Anderson impurity model [14]. In our approach exchange integrals are calculated exactly making use of atomic Hartree-Fock wavefunctions and angular momentum coupling technique and DOS are calculated making use of DFT package. Since present DFT packages use ultrasoft atomic potentials, the Hubbard parameters used in DFT calculations should be estimated relative to average of configuration values. As the result Hubbard parameters become smaller in magnitude and can have negative sign. According to our calculations part of DOS of Mn3$d$-electrons is strongly hybridized with the valence band at Fermi level. On the other hand the larger part of $d$-DOS is separated from this part and has a maximum at -2.7 eV. This result is in agreement with experimental [9] and theoretical results [9,21].

ASCNOWLEDGEMENTS

The work was supported by Russian Foundation for Basic Research, project 15-03-05370

Table 1. Atomic Coulomb integrals, coefficient for ground term relative to the average term and Hubbard values for ground term relative to the average term.

| Type of integral | Integral (eV) | coefficient | Hubbard value (eV) | $\Delta T_c$ |
|---|---|---|---|---|
| $R^2(3d3d\|1/r_{12}\|3d3d)$ | 10,31 | -.3968 | -1,327[*)] | 145 K |
| $R^4(3d3d\|1/r_{12}\|3d3d)$ | 6,41 | -.3968 | | |
| $R^2(3d3s\|1/r_{12}\|3s3d)$ | 0,889 | -0.5 | -0,444 | 21 K |
| $R^1(3d4p\|1/r_{12}\|4p3d)$ | 0,391 | -.3333 | -0,196 | |
| $R^3(3d4p\|1/r_{12}\|4p3d)$ | 0,313 | -.2143 | | |
| Total $T_c$ | | | | 166 K |

*) Per one electron in $3d^5$ shell

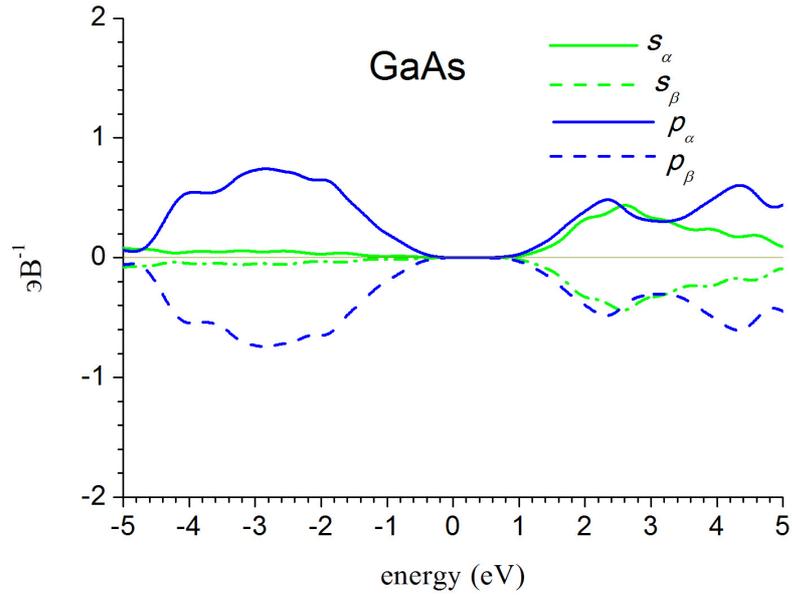

Figure 1. Theoretical spin-polarized densities of states of the host semiconductor GaAs.

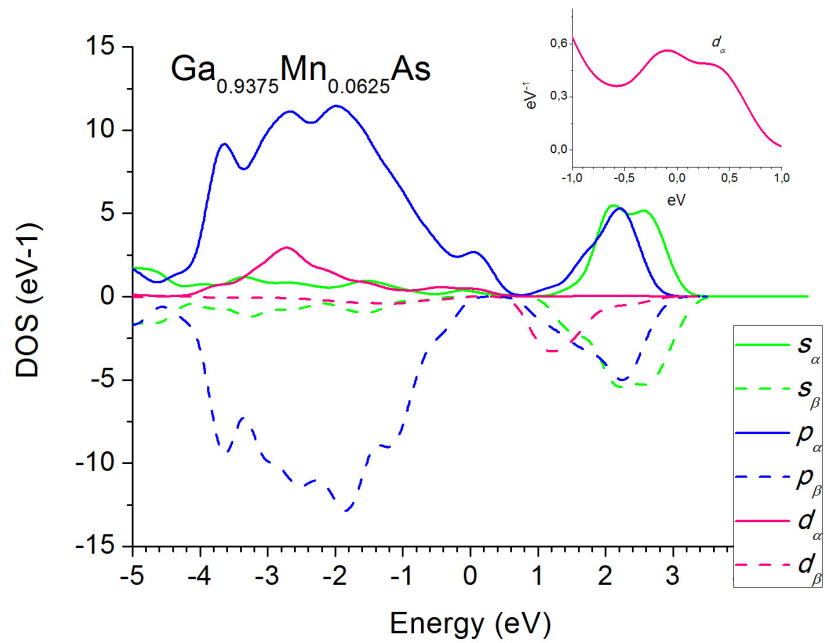

Figure 2. Theoretical spin-polarized densities of states of the dilute magnetic semiconductor $Ga_{0.9375}Mn_{0.0625}As$. DOS of *d*-electrons near $E_F$ are shown in the inset.

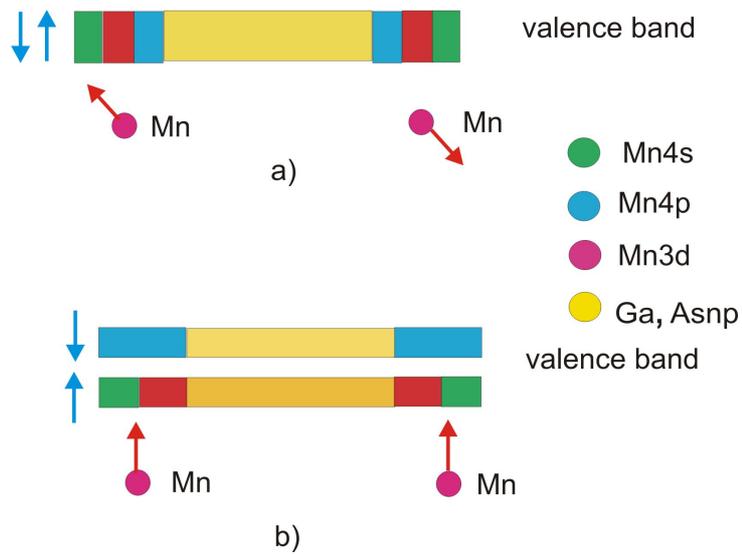

Figure 3. Model of the exchange energy gain due to redistribution of population of *s*-, *p*- and *d*-electrons between spin-up and spin-down subbands at magnetic centers (Mn)